\documentclass[usenatbib]{mnras}
\usepackage{amssymb}

\usepackage{newtxmath}

\usepackage[T1]{fontenc}
\usepackage[normalem]{ulem}

\DeclareRobustCommand{\VAN}[3]{#2}
\let\VANthebibliography\thebibliography
\def\thebibliography{\DeclareRobustCommand{\VAN}[3]{##3}\VANthebibliography}

\usepackage{graphicx}	
\usepackage{amsmath}	
\usepackage{amssymb}	

\title[Phase spirals in cosmological simulations]{Phase spirals in cosmological simulations of Milky Way-size galaxies}

\author[B. García-Conde et al.]{
B. García-Conde,$^{1}$\thanks{E-mail: begona01@ucm.es}
S. Roca-Fàbrega,$^{1}$
T. Antoja$^{2}$, P. Ramos$^{3}$, and 
O. Valenzuela$^{4}$
\\
$^{1}$Dpto. Física de la Tierra y Astrofísica, Universidad Complutense de Madrid, Madrid, Spain\\
$^{2}$Institut de Ciènces del Cosmos, Universitat de Barcelona, Barcelona 08028, Spain\\
$^{3}$ Observatoire astronomique de Strasbourg, Université de Strasbourg, Strasbourg 67000, France \\
$^{4}$ Universidad Nacional Autónoma de México, Instituto de Astronomía, AP 70-264, CDMX 04510, México}

\date{Accepted 2021 November 22. Received 2021 November 22; in original form 2021 September 17}

\pubyear{2021}

\begin{document}
\label{firstpage}
\pagerange{\pageref{firstpage}--\pageref{lastpage}}
\maketitle


\begin{abstract}
We study the vertical perturbations in the galactic disc of the Milky Way-size high-resolution hydrodynamical cosmological simulation named GARROTXA. We detect phase spirals in the vertical projection $Z- V_{Z}$ of disc's stellar particles for the first time in this type of simulations. Qualitatively similar structures were detected in the recent Gaia data, and their origin is still under study. In our model the spiral-like structures in the phase space are present in a wide range of times and locations across the disc. 
By accounting for an evolving mix of stellar populations, we observe that, as seen in the data, the phase spirals are better observed in the range of younger-intermediate star particles. 
We measure the intensity of the spiral with a Fourier decomposition and find that these structures appear stronger near satellite pericenters. Current dynamical models of the phase spiral considering a single perturber required a mass at least of the order of 10$^{10}$ M$_\odot$, but all three of our satellites have masses of the order of $\sim$10$^8$ M$_\odot$. We suggest that there are other mechanisms at play which appear naturally in our model such as the physics of gas, collective effect of multiple perturbers, and a dynamically cold population that is continuously renovated by the star formation
Complementing collisionless isolated N-body models with the use of fully-cosmological simulations with enough resolution can provide new insights into the nature/origin of the phase spiral.

\end{abstract}




\section{Introduction}\label{sec:intro}

One of the current goals in astrophysics is to understand how disc galaxies, and in particular our own Milky Way, form and evolve and to identify the processes that gave them shape. 
Recently several studies showed that our Galaxy's disc is highly perturbed \citep[e.g.]{widrow2012galactoseismology,williams2013wobbly, antoja18, ramos2018riding,Antoja2021}, which is especially evident with Gaia data \citep{gaia2018gaia}. 
The effects of these perturbations can be observed, for instance, as a one-armed spiral in the vertical phase space ($Z-V_{Z}$) of the Solar Neighbourhood, also when weighted by the rotational, $V_{\phi}$, or radial, $V_{R}$, velocities \citep{antoja18}. That study suggested that the phase spiral is a phase mixing signature after a perturbation. Further modelling has confirmed that the phase spiral can be a consequence of the Sagittarius dwarf galaxy tidal interaction \citep{binney2018origin, laporte2019footprints, Li2020, Bland-Hawthorn2021, hunt2021resolving, gandhi2021snails, widmark2021weighing}. Indeed, the estimated time of the perturbation roughly coincides with some of the previous pericenters of Sagittarius, estimated to be between 200 and 1000\,Myr ago \citep{law2010sagittarius, purcell2011sagittarius, Vasiliev2020}.
Nonetheless, other authors have presented alternative hypothesis where the bar buckling \citep{khoperskov2019buckling} or the halo substructure \citep{Darling2019} generate bending waves that can also result in phase spirals.

The studies mentioned above consist of analytical models using the impulse approximation or isolated N-Body simulations considering a single perturber aimed at reproducing the formation of the phase spiral in simplified and controlled conditions. Although a single satellite galaxy could be the dominant cause of the phase spiral, it is not straightforward to infer how the addition of processes such as the gas effects or multiple perturbers and their derived collective effects \citep{Weinberg1989} will influence the vertical perturbations.

In this work, we go one step further by studying the phase spiral in a high-resolution cosmological (N-body + hydrodynamics) zoom-in simulation for the first time. At the cost of having lower resolution and a larger analysis complexity than pure N-body galaxy models, we gain on realism by including most of the known physical mechanisms that shape galaxies: gas, star formation, formation 
of a galactic bar, and multiple satellites perturbing the disc. In addition, we are able to study the full history of the galaxy, all in a self-consistent manner. 

The detection of disc phase spirals in cosmological simulations was a challenge due its complexity, lack of tuning, and especially the resolution limits of these type of simulations, which may lead to a blurring of the phase space structures due to numerical diffusion 
\citep{sellwood2012spiral, colombi2021phase}.  
However, our model, which is described in Sect \ref{sec:methodology},
 allows us to explore the disc phase space at a resolution of $\sim$100~pc. Here we report the detection of resolved phase spirals in a cosmological simulation for the first time (Sect \ref{sec:spirals}) which appear to be common 
 throughout the evolution of the galaxy. 
 We  study its correlation with the pericenter passages of the three main satellites, as well as with the star formation history (Sect.~\ref{sec:satellites}). We conclude in Sect.~\ref{sec:discussion} discussing the possible origin of the phase spiral and describing the new avenues opened by analysing kinematics of galaxies in a fully cosmological context.

\section{Methodology}\label{sec:methodology}

  
The GARROTXA simulations \citep{garrotxa} are a set of zoom-in cosmological simulations of Milky Way-Mass galaxies. The spatial resolution of these models is of 100 pc, with a minimum dark matter particle mass of $10^5$ $M_{\odot}$, a mean stellar mass of $\sim4\times10^3$ $M_{\odot}$, and a minimum time-step of $10^3$ yr. This resolution allows us to resolve the disc scale length and height. The model we analyse here was generated using the  hydrodinamical version of the ART code \citep{kr97ART} and contains two galactic systems: a MW-mass galaxy, and an Andromeda-mass companion at 1~Mpc at $z=0$. 
We focus on the MW-mass system \citep[model G.322 in ][]{garrotxa} that has a virial radius $R_{\rm{v}}$ of 160 kpc,  computed following \citet{bryan1998statistical} where the spherical collapse model is used to determine the virial overdensity as a function of the redshift $\Delta_{\rm{v}}(z)$ and taking this value to be 333 (i.e., $R_{\rm{333}}$=$R_{\rm{v}}$, hereafter). The enclosed mass within this radius is $M_{\rm{v}}$ of $6.5\times 10^{11}$ M$_{\odot}$. \citet{garrotxa} give more details on the spatial, mass and temporal resolution, cosmology (see their Table 1), galaxy's rotation curve, and on the disc’s surface density profile, which agree well with observations for the MW. The disc has a persistent two exponential profile (see their figures 8 and 9) with a boxy-peanut galactic bar that has a secular origin (see their section 3.1.1 and Figure 2), and shows a slowly decreasing pattern speed from ~50 km/s/kpc at z=1 to $\sim$ 40 km/s/kpc at z=0 and a length evolving from $\sim$3 to $\sim$5 kpc.
We focus on the late times of the simulation which we re-simulated for this work, saving a snapshot every $10-50$ Myr of evolution.

The galaxy suffered its last major merger at $z=1.5$, and at $z=0$ it has a complex environment that includes multiple tidal streams and satellites. Obtaining the mass of satellites is not trivial in simulations, nor in observations. Once the satellite enters the densest regions of the host galaxy dark matter halo the outer parts of its own halo are quickly disrupted. Also, inside the virial/tidal radius of the satellite there is a mixture of the dark matter particles that are bound to the host with the ones that are to the satellite. Therefore, the dynamical mass does not reflect the total mass enclosed within a sphere centered on the satellite's center of mass at each time, 
and it is not fully correct to fit a NFW density profile either. 
Consequently, we decided to obtain the satellites' mass at $z\sim$2, when satellites are still outside $R_{v}$ of the host galaxy, using two independent techniques: computing the total mass enclosed in its own $R_{v}$, and via abundance matching \citep{rodriguez2017constraining, behroozi2010comprehensive}. In Table \ref{tab:satellites} we show both results in the column labeled as $M_{inf}$. Results from the virial approximation (first value) are systematically lower than the ones from abundance matching (second). This result is not surprising as dwarf satellites may suffer tidally induced star formation and dark matter stripping well before entering the host's virial radius \citep{Guo2020,Jackson2021}, both acting against the theoretical M$_*$/M$_v$ relation used in the abundance matching approach.

From 6 to 0 Gyr in lookback time we defined their radii as their tidal radius at apocentre (where it is easier to calculate) and computed the mass within. This radius is updated at each apocentre to account for the mass loss at pericentre. These values are shown in the second column of the table.

\begin{table}
	\centering
	\caption{
	Properties of the three largest satellite galaxies in the simulation. From left to the right: total mass at first infall (we give two different values corresponding to the mass inside the virial radius $M_{v}$ and the one from abundance matching at $z\sim2$,  respectively); tidal radius, total mass, stellar mass, and orbit inclination with respect to the galactic plane with the two values being at 6 and 0 Gyr lookback time.}
	\label{tab:satellites}
	\begin{tabular}{lccccc} 
		\hline
		&$M_{\rm inf}$ & $R_{\rm t}$ & $M_{\rm T}$ & $M_*$ & $i$ \\
Satellite & [10$^{10}$~M$_{\odot}$] &[kpc] & [10$^{8}$~M$_{\odot}$] & [10$^{8}$~M$_{\odot}$] & [deg]\\
		\hline
		Arania& 1.3 $\vert$ 6.6 & 3.0-2.0 & 5.4-2.5 & 2.30-1.50 & 103-120 \\
		Grillo&0.1 $\vert$ 0.9 & 1.4-1.0  & 1.8-1.1  & 0.15-0.07 & 34-23   \\
		Mosquito&0.2 $\vert$ 2.5& 1.0-1.0 &  0.7-0.6 &   0.04-0.03 & 75-97  \\ 
  
		\hline
	\end{tabular}
\end{table}

To read and analyse the simulation we adapt the yt-based AGORA toolkit \citep{yt}, used by the AGORA community \citep{RocaFabrega2021}. For each snapshot, we localize the galactic center in the cosmological box and align the disc with the $Z$ axis defined by the angular momentum of stellar particles. 
 Our method consists in a two step alignment. Firstly, we use yt to calculate the angular momentum vector in a sphere containing the disc (0.1 R$_v$), with stellar particles less than 5 Gyr old, that dominate the angular momentum of the thin disc.

 We align the $Z$ axis with this vector. After that, we take all stellar data and apply a second alignment, this time taking a cylinder with 15 kpc of radius and recalculating $L$ with all stars within. 
We use galactocentric cylindrical coordinates $\phi$, $R$ and $Z$, with $\phi$ and $V_{\phi}$ being negative in the direction of rotation.

\section {Phase spirals through space and time}\label{sec:spirals}

In this section, we analyse the vertical projection of the phase space of the disc in GARROTXA  at different times,  different volumes, and for different populations.

\begin{figure}   

	\includegraphics[width=\columnwidth]{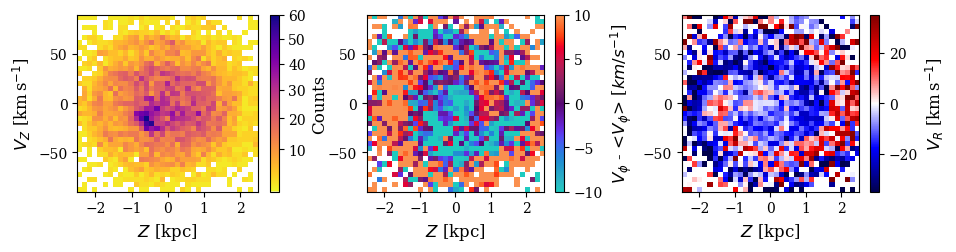}
    \caption{Phase space spirals ($Z- V_{Z}$) observed in the 0-5 Gyr  stellar particles' age population of the GARROTXA simulation at 2.74 Gyr in lookback time, that is 1.1 Gyr after the pericenter of the most massive satellite (Arania) and 0.7 Gyr after the second most massive (Grillo). From left to right: a two-dimensional histogram with no weighting, $V_{\phi}$ weighted and $V_{R}$ weighted.}

    \label{fig:example}
\end{figure}

Initially, we focus our analysis on stellar particles with ages between 0 and 5 Gyr (but see below an exploration with age). 
We {first} take stellar particles with galactocentric radii from 10 to 12 kpc and vertical position $|Z|< 2.5$ kpc, then divide this annulus in twelve adjacent sectors spanning 30 $\deg$ in $\phi$, and we follow their temporal evolution from lookback time of 6 to 0 Gyr.
We present a first example of the phase-space spirals ($Z- V_{Z}$) in Fig.~\ref{fig:example}. In this figure we display a single volume for a snapshot that is at 1.1 Gyr after the pericenter of the most massive satellite Arania. We show the vertical projection of phase space in the range of [-2.5, 2.5] kpc in the vertical position axis and [-80,80] km/s in the velocity axis with a 35 x 35 binning, in density (first column), weighted by the velocity $V_\phi$ (second column), and by $V_{R}$ (third column). Like in most of the previous studies using N-Body simulations, we also note that the agreement with the phase spiral in \emph{Gaia} is only qualitative, since we have much larger volumes and a much smaller number of particles. We see that the phase spirals are clearer when coloured by velocities, especially $V_\phi$, as in observations \citep{antoja18}. However, we do not see as many wraps of winded phase spirals as in the \emph{Gaia}  data. Of course, this can be due in part to the numerical diffusion. We want to emphasise, though, that our objective is not to reproduce the \emph{Gaia} phase spiral in its details, which is a really challenging task in a cosmological simulation, but to detect qualitatively similar phase spirals originating in a more complex model than the ones seen so far.

\begin{figure*}   

	\includegraphics[width=0.98\textwidth]{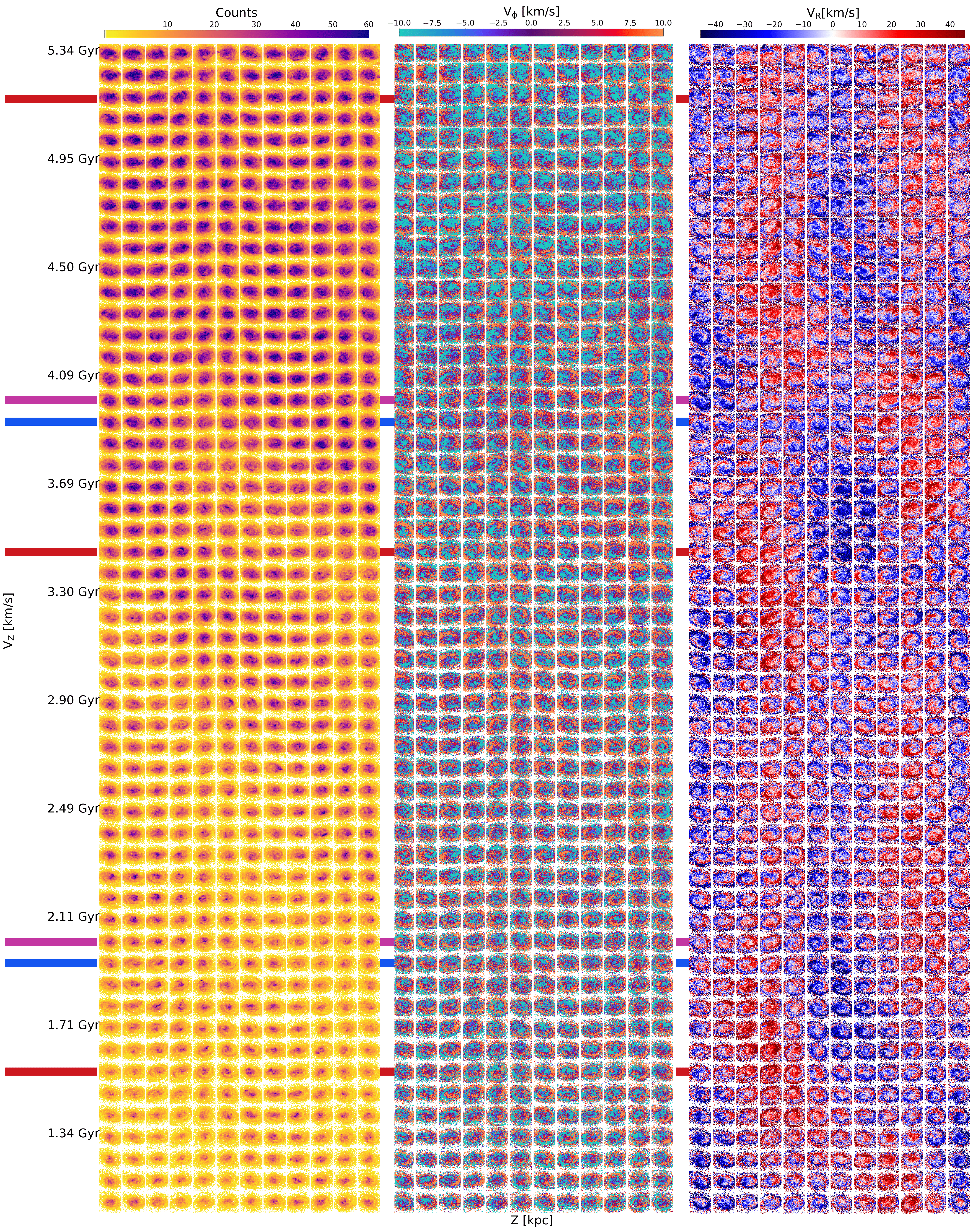}
    \caption{Phase spirals in the vertical phase space at different azimuths and times of the GARROTXA simulation. We consider particles at radius between 10 and 12 kpc and with ages younger than 5 Gyr. From top to bottom we show the evolution between 5.3 - 1 Gyr in lookback time. From left to right, the first set of 12 columns show the vertical phase space in 12 azimuthal equispaced regions ($30\deg$). The next 12 columns show the $V_{\phi} - <V_{\phi}>$ weighted maps, where  $<V_{\phi}>$ is the mean for each region. The last set shows the $V_{R}$ weighted maps. The horizontal thick lines indicate the pericenters of the three main satellites (Arania in blue, Grillo in red and Mosquito in magenta).}

    \label{fig:spiral}
\end{figure*}

 In Fig.~\ref{fig:spiral} we show the vertical phase space density in each of these  12 regions defined above but for all snapshots, between 6 and 0 Gyr, in density (first column), weighted by the velocity $V_\phi$ (second column), and by $V_{R}$ (third column). Typically, in each region we find about 5000 to 19000 stellar particles. In all panels we see a non-uniform distribution, with most of them having a spiral shape.
In fact, for the first snapshots that we study there is already a certain degree of spirality. We note that there are previous passages of the satellites before 5.5 Gyr in lookback time, being the first infall of these satellites at $\sim$11-12 Gyr. The formation of the stellar thin disk in our model is enhanced by these first pericenters, and, once formed, it remains almost permanently disturbed by the many following interactions with the satellites. However, we can see that the phase spiral becomes more perceptible around the 4 Gyr. At later times, the vertical phase space contains also thin short phase spirals, separated from the main distribution. Some of them seem to be formed by particles of similar ages, and therefore they could be phase mixing structures from particular star forming groups.

If we analyse the panels as a whole we observe retrograde diagonal patterns that go from top-right to bottom- left, especially in the density and weighted by $V_{\phi}$. In the thinner time step version of the figure, thinner pro-grade bands are also observed, as seen in Fig \ref{fig:SFH} as fourierograms.
There are also bands in Fig.~12 of \citet{Bland-Hawthorn2021}, who presented a similar figure, but they have different slope and a clear $m=2$ mode consistent with their bending wave model (the $Z$-$V_Z$ distribution repeats twice azimuthally) instead of an $m=1$ like in our case. These patterns indicate that the perturbation has an angular dependence, and it  moves through the disc. These bands are better observed in Fig.~\ref{fig:SFH} and we come back to this later on.

\begin{figure}
	\includegraphics[width=1\columnwidth]{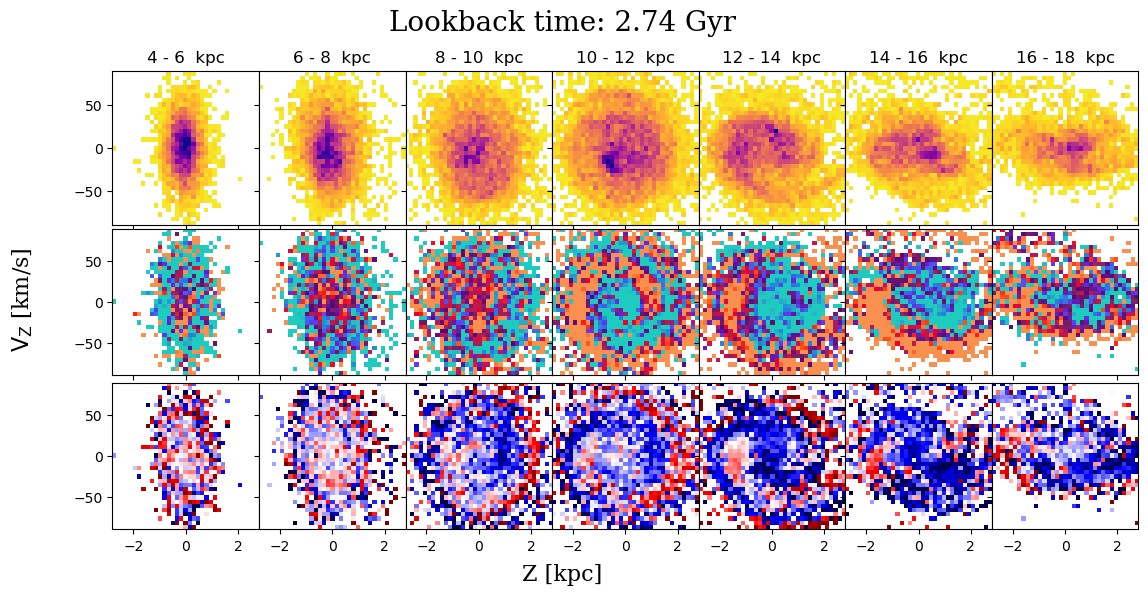}
    \caption{Vertical phase space distributions of stellar particles with age of $0-5$ Gyr at seven different radial bins (see top labels) and at a fixed azimuthal direction of 330$^{\circ}$ inside the disc, from the snapshot at 2.74~Gyr of lookback time. We show the phase space density (top), $V_{\phi} - <V_{\phi}>$ weighted (middle), and $V_{R}$ weighted (bottom).  $<V_{\phi}>$ is the mean azimuthal velocity at each region.}
 \label{fig:radii}
\end{figure}

In Fig.~\ref{fig:radii} we explore the vertical phase space structures as a function of radius for a single azimuth. We show the phase space density (top), $V_{\phi}$-weighted (middle), and $V_{R}$-weighted (bottom). We see that the global distribution changes from elongated in the velocity axis at small radii to elongated in $Z$ in the outer disc. This  has been detected in the Gaia data \citep{laporte2019footprints} and is a consequence of the smaller restoring vertical force in the outer parts of the disc. The phase spiral becomes detectable at radii larger than 6 kpc and is present up to very large radius, showing that they are not exclusive of the range of 10-12 kpc chosen before. The short vertical extent of the distribution at inner radii, combined with the limitations in resolution, may be hampering our ability to see clear structure there. 

\begin{figure}
	\includegraphics[width=1\columnwidth]{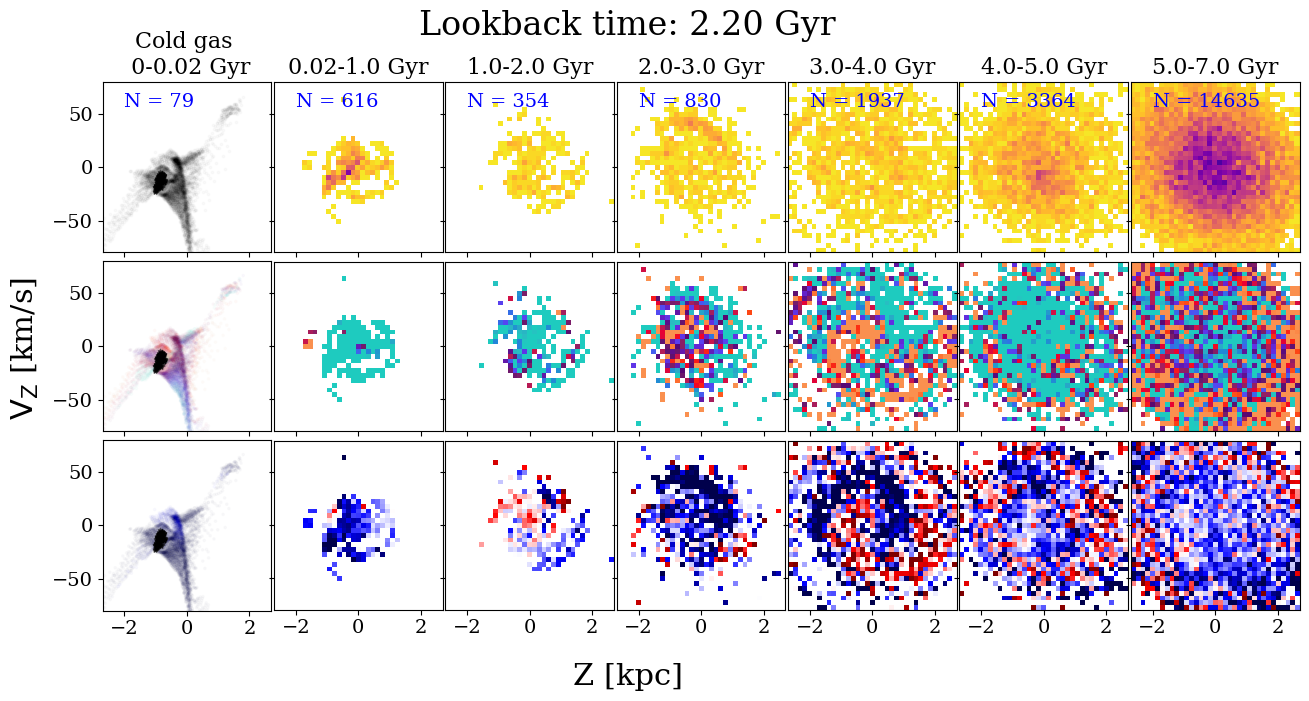}
    \caption{Vertical phase space of seven age populations within a region with fixed azimuth (150$^{\circ}$) and distance to the galactic center (10-12 kpc), at lookback time of 2.2 Gyr. From top  to  bottom we show: density of stars; $V_{\phi} - <V_{\phi}>$ weighted; $V_{R}$ weighted. Additionally, the first column shows the cold gas and the newborn stellar particles as black dots.}
 \label{fig:edades}
\end{figure}

We show in Fig.~\ref{fig:edades} seven stellar populations with increasing age, starting with the cold gas (< 8000 K) and newborn stellar particles. 
We see that the global phase spiral is more pronounced in stellar particles with ages of about $1-5$~Gyr confirming the age analysis in \citet{Tian2018} and \citet{bland2019galah}. Although there are some hints of structure in the older groups, it definitively fades out for particles older than 5 Gyr. This is due to younger stars being dynamically colder, thus reflecting more prominently the effects of perturbing phenomena. Also, as described in \citet{Li2020}, groups of older stars whose orbits are kinematically hot have a larger range of vertical frequencies, which may blur the phase spiral. We note here that the spiral pattern observed in the $V_{\phi}$-weighted maps (central row) differs from one age population to another (e.g. the $2-3$ Gyr group vs. the $3-4$~Gyr one). This result suggests that different stellar populations may have been perturbed and/or phase-mixed differently.  
Finally, we see that the very young stellar particles (less than 1~Gyr) are found in groups that do not fully cover the phase space but present some sort of spiralility. In fact, we see in the first column that the cold gas appears to be distributed in non-isotropic phase space patterns and the newborn stellar particles (black dots) are not born close to $Z\sim0$ and $V_Z\sim0$. This will lead to subsequent phase mixing of the young populations which could create the thin spirals in the vertical phase space projection. 

Above we show qualitatively that we detect phase spirals at multiple times and locations, but it is not trivial to distinguish the moment of their appearance. Here we present a method to quantitatively discern the emergence and development of said phase spirals. We do a Fourier analysis of the vertical phase space of the same disc regions and young and intermediate age population ($0-5$~Gyr) as in Fig.~\ref{fig:spiral}. We normalize $V_{Z}$ with the dispersion of $Z$ so that both axis have the same scale. 
In this way, the distribution of the phase space have a circular shape, which can be divided in annular bins to proceed with the Fourier modes calculation at each one.
A phase spiral will present a high amplitude of the m=1 mode in the $Z$-$V_Z$ space (or m=2 mode if two-armed) and will have a phase of maximum amplitude ($\phi_{\rm{max}}$) varying with distance to the $(Z, V_Z)=(0,0)$ point. We require that the m=1 amplitude relative to m=0 is 1.1 times larger than for higher modes (from m=3 to m=6), and that $\phi_{\rm{max}}$  has a standard deviation of more than 30 deg across the annular bins. If these conditions are not fulfilled, we set the amplitude to 0, which will reduce the number of false positives due to noise or to the presence of a bi-modal distribution without angular dependence. We exclude the two first and last bins, since those are the ones which often are more prone to generate noise.

The result of this analysis is the map of the temporal and spatial evolution of the strength of the phase spiral in density and also $V_{\phi}$ and $V_{R}$-weighted (first three panels of Fig.~\ref{fig:SFH}). In these panels we see how spirals appear and propagate in time (right to left) and azimuth (bottom to top), following diagonal patterns consistent to the ones seen in the extended version of Fig. \ref{fig:spiral}.
We estimate that the azimuth propagation of phase spirals with high Fourier amplitude has a period of about 300 Myr, which is compatible with the rotation period at the range of 10-12 kpc in radius (250-320 Myr).

These bands are of unknown origin but we remark that they are not equivalent to the bands seen in Fig.~12 of \citet{Bland-Hawthorn2021} where these bands correspond to volumes that are re-aligned with the impact site of the perturber at outer radius. Our bands have a higher frequency than the re-encounters with impact sites. In fact, in our model most of the pericentres occur outside the disc region and even if we consider those points, we get re-encounter times of about 400 Myr or larger. We note that the differences between our bands and those of \citet{Bland-Hawthorn2021} are not surprising since the models that we are comparing here are significantly different (our satellites impact points are outside of the disc in most of the cases, the interactions are not as impulsive as in their model, and we may have collective effects due to the presence of many satellites at once, with pericenters at very similar times, as discussed in Sect.\ref{sec:discussion}). Nonetheless, the period of these bands must be related to the physics of such phenomena and are worth investigating in the future.

Globally, the phase spirals are significantly more intense at certain times, which was already noticed in Fig.~\ref{fig:spiral}, but is now quantified using our Fourier technique. 
The first three rows of Fig.\ref{fig:SFH} present interesting differences. For example, the amplitudes of the $V_{\phi}$ weighted and $V_{R}$ weighted spirals (second and third rows, respectively) are higher around 4~Gyr while the ones for the density (first row) are higher around 2~Gyr. Moreover, the duration of high intensity phases is different, with the $V_{R}$-weighted coefficients decaying faster than for $V_{\phi}$. However, at recent times (< 2 Gyr), there are progressively less particles and the distribution is more compact in the $Z$-$V_Z$ space, which may have some effect on our strength estimator applied to different quantities (density or velocity-weighted).

\section{Satellites and Star Formation}\label{sec:satellites}

\begin{figure}
	\includegraphics[width=1\columnwidth]{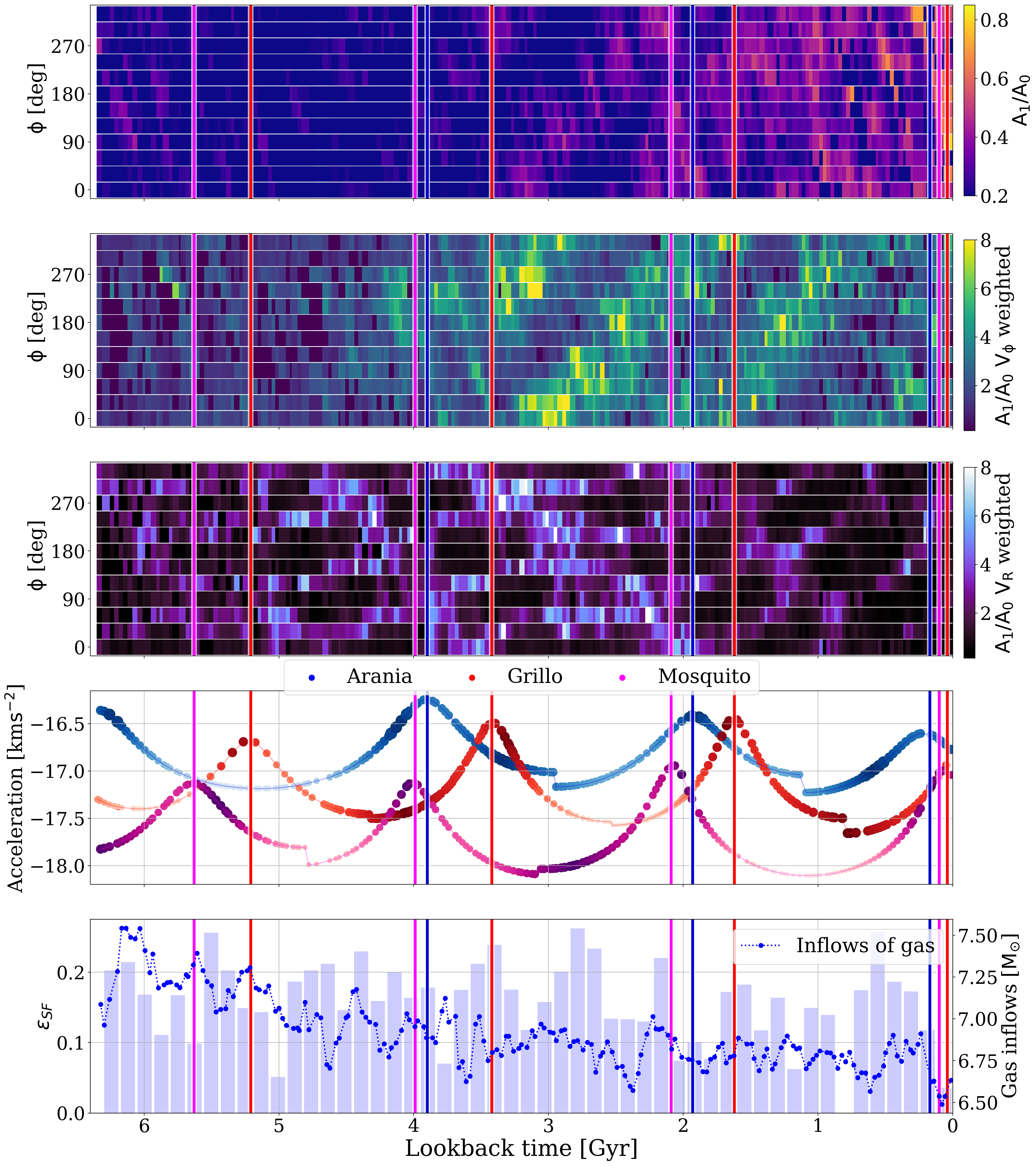}
    \caption{Amplitude of the phase spiral with time and relation with satellite orbits and star formation. First three panels: m=1 Fourier amplitude in the Z-$V_Z$ space, for the same regions and ages as in Fig. \ref{fig:spiral}, computed for the density, $V_{\phi}$-weighted, and $V_{R}$-weighted, respectively. Fourth panel: acceleration onto the disc by the main satellites with darker colours indicating a lower vertical distance above/below the disc plane. Here the small discontinuities are artificially caused by our re-calculation of the tidal radius at apocentres. Bottom panel: star formation efficiency (histogram) and gas inflow histories (blue dots and line). Coloured vertical lines indicate the times of pericentres.}\label{fig:SFH}%
\end{figure}

In this section we analyse the relation of the phase spiral with the three main satellites in our model. We first characterized their properties and orbits (Tab.~\ref{tab:satellites}). We then computed the mean acceleration onto the disc applied by each satellite as a function of time (fourth panel of Fig.~\ref{fig:SFH}).

The satellites present different levels of impulsiveness. They also have different orbital inclinations, with Arania and Mosquito having almost polar orbits and Grillo, a more planar one. Also, some pericentres coincide with the point along the orbit of smallest altitude below/above the plane, as indicated by the darker colours in Fig. \ref{fig:SFH} (e.g., the pericenters at 2~Gyr). Some other pericenters occur while the influence of previous ones must still be ongoing (e.g. Grillo's pericenters at 3.5~Gyr happens after the recent pericenters at 4~Gyr of Arania and Mosquito). Interestingly, we also see that, although the main satellites have different infall times, their pericenters tend to synchronize, occurring almost simultaneously by $z = 0$. 
 
Comparing the fourth panel with the results from the Fourier analysis (three top panels) we see low amplitudes of the phase spiral at initial times but a clear coincidence between the presence of well defined phase spirals (warmer colors) and the pericenters (coloured vertical lines), for example after lookback times of 4 and 2 Gyr. All the complexity in the history of external perturbations described above might be the cause of the differences among the first three panels of Fig.~\ref{fig:SFH}. 
Although as mentioned previously, some differences might be caused by the estimator used, some others,  might be telling us about aspects of the perturber's orbits beyond merely pericentre times and masses, a characteristic that would be worth exploring further.

We have the advantage of having multiple pericenters (of different satellites) that appear naturally in our simulation. This, combined with the fact that in our model dynamically cold stars are being formed all along the evolution of the galaxy, results in different stellar populations responding differently to the new perturbations, which is something not captured by isolated models without star formation and gas. Moreover, the maximum strength of the gravitational pull of our satellites remains rather constant with time, in contrast with an interaction with a Sagittarius-like system, where every new pericentre induces a larger kick in velocity than the previous one, making it easier to “overwrite” the existing phase-space substructure. In any case, although we see a decrease of the signal after the 4 Gyr rise and a small increase/stabilization at about the time of the next pericenter, we can not confirm if the “reset” proposed by \citet{laporte2019footprints} and \citet{Bland-Hawthorn2021} occurs in our model after each pericenter.

For the sake of completeness, we also analysed the evolution of the gas and star formation in the disc. In the last panel of Fig. \ref{fig:SFH} we show the cold gas inflow rate (blue curve) defined as the amount of gas that penetrates a thin shell of $0.5$\,kpc at a radius of $0.15R_{\rm{v}}$
with a significant negative radial velocity ($< -50$ km/s). We also show the star formation efficiency ($\epsilon_{SF}$, blue histogram) computed as the fraction of stellar mass formed since the previous snapshot to the mass of gas available in the disc. This gas meets the star forming criteria of temperature ($T < 8000$ K) and density ($nH > 1$) just as in \citet{garrotxa}. With this panel we confirm that satellite pericenters not only perturb the disc's dynamics but also boost its star formation efficiency, with some coinciding with an increase in the cold gas inflow as well  (e.g. at 2 Gyr).

\section{Discussion and conclusions}\label{sec:discussion}

We report, for the first time, the detection of phase spirals in a realistic zoom-in cosmological model of a Milky Way-like system. These spirals are present throughout the last several Gyr of the evolution, suggesting that this phenomenon might be common in the life of certain galaxies. The spirals are more notable for younger to intermediate-age stars and are especially prominent near pericenter passages of the three main galaxy satellites. These passages coincide with the time when we observe star formation enhancements, an effect that has already been directly related to accretion events \citep[see ][for a discussion on the particular case of the Milky Way]{RuizLara2020}.

Most of the recent studies that modelled the phase spiral considered Sagittarius as the main culprit and used a heavy Sagittarius dwarf galaxy ($\gtrsim 10^{10}$ M$_{\odot}$ at the time of impact. For example, in \citet{binney2018origin} only a massive perturber was able to generate prominent phase spiral signals. Similarly, the masses considered in \citet{laporte2019footprints} and \citet{Bland-Hawthorn2021} are $6\cdot10^{10}$ and $2\cdot10^{10}$ M$_{\odot}$, respectively. Even more, although their work is based in a simplified one dimensional model, \citet{Bennett2021} recently claimed that even a heavy Sagittarius would not be able to generate the phase spirals seen in Gaia DR2, invoking the need for a combination of other effects. By contrast, our heaviest satellite (Arania), which has a mass at first infall of $\sim10^{10} M_{\odot}$, ends up with $\sim10^8$ $M_{\odot}$ after several pericenters. These values are in approximate agreement with the initial mass of Sagittarius (e.g. $\sim 10^{10}$ M$_{\odot}$ in \citealt{NiedersteOstholt2010}) and with  some recent empirical derivations of the total Sagittarius mass  at current times (e.g. $\sim4\times10^{8}$ M$_{\odot}$ in \citealt{Vasiliev2020}). However, the pericenter distance of our heaviest satellite is larger than the ones of recent Sagittarius pericenters, and thus the effects of our perturbers should be far less strong. In conclusion, our satellites seem to belong to the low mass regime, yet we do observe phase spirals and a correlation between pericenters and the strength of the phase spirals.

This can be attributed to the many extra ingredients that our model has with respect to previous models (e.g., the evolution in a cosmological framework, hydrodynamical processes, disc secular evolution) that can play a combined role in shaping the phase space. For example, we note that in our model pericenters get synchronized over time, and that there is a combined mass of up to $\sim10^9$ $M_{\odot}$ by the time they start affecting the disc. Collective effects such as the wake induced in the halo by infalling satellites have been shown recently to be of large importance \citep{Weinberg1998,Conroy2021} and in particular, coeval infalls can lead to the generation of a collective effect with strong impact on the disc kinematics (Trelles et al. in prep.). Other possible mechanisms that could be expected to produce perturbations in the phase space are: highly anisotropic distribution of dark matter left over from the satellites, multiple dark subhalos, misalignment between disc and halo, resonances, and the presence of non-axisymetric structures in the disc. In particular, the role of the internal structures such as the bar and spiral arms needs to be evaluated in detail since these phenomena can trigger vertical perturbations as suggested in \citet{khoperskov2019buckling}. Nonetheless, in our model we do not see signatures of a bar buckling in the studied interval of time.

Another important aspect that no other model of the phase spiral included before is the gas. The in-falling cold gas and the newborn stars can keep the disc kinematically cold and thus more unstable to internal and external perturbations. Additionally, as is the case of our model, the gas might be vertically perturbed and the young stellar particles could inherit its dynamical properties. We have observed the appearance of thin segments of phase spirals which seem to be connected to phase mixing of star formation complexes born from this perturbed gas. At this stage, however, it is not clear how these thin spirals are related to the global phase spirals. Interestingly, star forming regions organized in certain vertical patterns \citep{Alves2020} 
and global vertical disturbances in the gas \citep{Lallement2019} have been observed in the Milky Way.

We can still not claim a perfect match between the phase spirals observed in the Gaia data and the ones found in our model, yet it was not in the scope of this paper to find the same structures as the ones observed. Nevertheless, using simulations like the ones analysed here allows us to study the dynamical processes that give rise to the phase space spiral (or any other phase space structure, for that matter) in a much more realistic scenario for a MW-like galaxy. Moreover, the possibility of connecting local dynamical phenomena with global perturbations from satellites, gas behaviour and star formation processes in the same model in the context of a cosmological simulation is definitively a promising future avenue of work that we open with this paper.

\section*{Acknowledgements}

The authors wish to thank the anonymous Referee for her/his com- ments and suggestions that improved this work. BGC and SRF work has been supported by the Madrid Government (Comunidad de Madrid-Spain) under the Multiannual Agreement with Complutense University in the line Program to Stimulate Research for Young Doctors in the context of the V PRICIT. They also acknowledge financial support from the Spanish Ministry of Economy and Competitiveness (MINECO) under grant number AYA2016-75808-R, AYA2017-90589-REDT, RTI2018-096188-B-I00 and S2018/NMT-429, and from the CAM-UCM under grant number PR65/19-22462. SRF acknowledges support from a Spanish postdoctoral fellowship, under grant number 2017-T2/TIC-5592.  TA acknowledges the grant RYC2018-025968-I funded by MCIN/AEI/10.13039/501100011033 and by ``ESF Investing in your future''. This work was (partially) funded by the Spanish MICIN/AEI/10.13039/501100011033 and by ``ERDF A way of making Europe'' by the ``European Union'' through grant RTI2018-095076-B-C21, and the Institute of Cosmos Sciences University of Barcelona (ICCUB, Unidad de Excelencia ’Mar\'{\i}a de Maeztu’) through grant CEX2019-000918-M.
PR acknowledges support by the Agence Nationale de la Recherche (ANR project SEGAL ANR-19-CE31-0017 and project ANR-18-CE31-0006) as well as from the European Research Council (ERC grant agreement No. 834148). Simulations were performed on the {\sc Miztli} supercomputer at the LANACAD, UNAM, within the research project LANCAD-UNAM-DGTIC-151.


\section*{Data availability}

The data underlying this article will be shared on reasonable request to the corresponding author.
All figures are available at \url{https://github.com/Bego-GarciaConde/cosmological-phase-spirals-figures}



\bibliographystyle{mnras}
\bibliography{main} 

\begin{thebibliography}{}
\makeatletter
\relax
\def\mn@urlcharsother{\let\do\@makeother \do\$\do\&\do\#\do\^\do\_\do\%\do\~}
\def\mn@doi{\begingroup\mn@urlcharsother \@ifnextchar [ {\mn@doi@}
  {\mn@doi@[]}}
\def\mn@doi@[#1]#2{\def\@tempa{#1}\ifx\@tempa\@empty \href
  {http://dx.doi.org/#2} {doi:#2}\else \href {http://dx.doi.org/#2} {#1}\fi
  \endgroup}
\def\mn@eprint#1#2{\mn@eprint@#1:#2::\@nil}
\def\mn@eprint@arXiv#1{\href {http://arxiv.org/abs/#1} {{\tt arXiv:#1}}}
\def\mn@eprint@dblp#1{\href {http://dblp.uni-trier.de/rec/bibtex/#1.xml}
  {dblp:#1}}
\def\mn@eprint@#1:#2:#3:#4\@nil{\def\@tempa {#1}\def\@tempb {#2}\def\@tempc
  {#3}\ifx \@tempc \@empty \let \@tempc \@tempb \let \@tempb \@tempa \fi \ifx
  \@tempb \@empty \def\@tempb {arXiv}\fi \@ifundefined
  {mn@eprint@\@tempb}{\@tempb:\@tempc}{\expandafter \expandafter \csname
  mn@eprint@\@tempb\endcsname \expandafter{\@tempc}}}

\bibitem[\protect\citeauthoryear{{Alves} et~al.,}{{Alves}
  et~al.}{2020}]{Alves2020}
{Alves} J.,  et~al., 2020, \mn@doi [\nat] {10.1038/s41586-019-1874-z}, \href
  {https://ui.adsabs.harvard.edu/abs/2020Natur.578..237A} {578, 237}

\bibitem[\protect\citeauthoryear{{Antoja} et~al.,}{{Antoja}
  et~al.}{2018}]{antoja18}
{Antoja} T.,  et~al., 2018, \mn@doi [\nat] {10.1038/s41586-018-0510-7}, \href
  {https://ui.adsabs.harvard.edu/abs/2018Natur.561..360A} {561, 360}

\bibitem[\protect\citeauthoryear{{Behroozi}, {Conroy}  \&
  {Wechsler}}{{Behroozi} et~al.}{2010}]{behroozi2010comprehensive}
{Behroozi} P.~S.,  {Conroy} C.,   {Wechsler} R.~H.,  2010, \mn@doi [\apj]
  {10.1088/0004-637X/717/1/379}, \href
  {https://ui.adsabs.harvard.edu/abs/2010ApJ...717..379B} {717, 379}

\bibitem[\protect\citeauthoryear{{Bennett} \& {Bovy}}{{Bennett} \&
  {Bovy}}{2021}]{Bennett2021}
{Bennett} M.,  {Bovy} J.,  2021, \mn@doi [\mnras] {10.1093/mnras/stab524},
  \href {https://ui.adsabs.harvard.edu/abs/2021MNRAS.503..376B} {503, 376}

\bibitem[\protect\citeauthoryear{{Binney} \& {Sch{\"o}nrich}}{{Binney} \&
  {Sch{\"o}nrich}}{2018}]{binney2018origin}
{Binney} J.,  {Sch{\"o}nrich} R.,  2018, \mn@doi [\mnras]
  {10.1093/mnras/sty2378}, \href
  {https://ui.adsabs.harvard.edu/abs/2018MNRAS.481.1501B} {481, 1501}

\bibitem[\protect\citeauthoryear{{Bland-Hawthorn} \&
  {Tepper-Garc{\'\i}a}}{{Bland-Hawthorn} \&
  {Tepper-Garc{\'\i}a}}{2021}]{Bland-Hawthorn2021}
{Bland-Hawthorn} J.,  {Tepper-Garc{\'\i}a} T.,  2021, \mn@doi [\mnras]
  {10.1093/mnras/stab704}, \href
  {https://ui.adsabs.harvard.edu/abs/2021MNRAS.tmp..722B} {}

\bibitem[\protect\citeauthoryear{{Bland-Hawthorn} et~al.,}{{Bland-Hawthorn}
  et~al.}{2019}]{bland2019galah}
{Bland-Hawthorn} J.,  et~al., 2019, \mn@doi [\mnras] {10.1093/mnras/stz217},
  \href {https://ui.adsabs.harvard.edu/abs/2019MNRAS.486.1167B} {486, 1167}

\bibitem[\protect\citeauthoryear{{Bryan} \& {Norman}}{{Bryan} \&
  {Norman}}{1998}]{bryan1998statistical}
{Bryan} G.~L.,  {Norman} M.~L.,  1998, \mn@doi [\apj] {10.1086/305262}, \href
  {https://ui.adsabs.harvard.edu/abs/1998ApJ...495...80B} {495, 80}

\bibitem[\protect\citeauthoryear{{Colombi}}{{Colombi}}{2021}]{colombi2021phase}
{Colombi} S.,  2021, \mn@doi [\aap] {10.1051/0004-6361/202039719}, \href
  {https://ui.adsabs.harvard.edu/abs/2021A&A...647A..66C} {647, A66}

\bibitem[\protect\citeauthoryear{{Conroy}, {Naidu}, {Garavito-Camargo},
  {Besla}, {Zaritsky}, {Bonaca}  \& {Johnson}}{{Conroy}
  et~al.}{2021}]{Conroy2021}
{Conroy} C.,  {Naidu} R.~P.,  {Garavito-Camargo} N.,  {Besla} G.,  {Zaritsky}
  D.,  {Bonaca} A.,   {Johnson} B.~D.,  2021, \mn@doi [\nat]
  {10.1038/s41586-021-03385-7}, \href
  {https://ui.adsabs.harvard.edu/abs/2021Natur.592..534C} {592, 534}

\bibitem[\protect\citeauthoryear{{Darling} \& {Widrow}}{{Darling} \&
  {Widrow}}{2019}]{Darling2019}
{Darling} K.,  {Widrow} L.~M.,  2019, \mn@doi [\mnras] {10.1093/mnras/sty3508},
  \href {https://ui.adsabs.harvard.edu/abs/2019MNRAS.484.1050D} {484, 1050}

\bibitem[\protect\citeauthoryear{{Gaia Collaboration} et~al.,}{{Gaia
  Collaboration} et~al.}{2018}]{gaia2018gaia}
{Gaia Collaboration} et~al., 2018, \mn@doi [\aap]
  {10.1051/0004-6361/201833051}, \href
  {https://ui.adsabs.harvard.edu/abs/2018A&A...616A...1G} {616, A1}

\bibitem[\protect\citeauthoryear{{Gaia Collaboration} et~al.,}{{Gaia
  Collaboration} et~al.}{2021}]{Antoja2021}
{Gaia Collaboration} et~al., 2021, \mn@doi [\aap]
  {10.1051/0004-6361/202039714}, \href
  {https://ui.adsabs.harvard.edu/abs/2021A&A...649A...8G} {649, A8}

\bibitem[\protect\citeauthoryear{{Gandhi}, {Johnston}, {Hunt}, {Price-Whelan},
  {Laporte}  \& {Hogg}}{{Gandhi} et~al.}{2021}]{gandhi2021snails}
{Gandhi} S.~S.,  {Johnston} K.~V.,  {Hunt} J. A.~S.,  {Price-Whelan} A.~M.,
  {Laporte} C. F.~P.,   {Hogg} D.~W.,  2021, arXiv e-prints, \href
  {https://ui.adsabs.harvard.edu/abs/2021arXiv210703562G} {p. arXiv:2107.03562}

\bibitem[\protect\citeauthoryear{{Guo} et~al.,}{{Guo} et~al.}{2020}]{Guo2020}
{Guo} Q.,  et~al., 2020, \mn@doi [Nature Astronomy]
  {10.1038/s41550-019-0930-9}, \href
  {https://ui.adsabs.harvard.edu/abs/2020NatAs...4..246G} {4, 246}

\bibitem[\protect\citeauthoryear{{Hunt}, {Stelea}, {Johnston}, {Gandhi},
  {Laporte}  \& {B{\'e}dorf}}{{Hunt} et~al.}{2021}]{hunt2021resolving}
{Hunt} J. A.~S.,  {Stelea} I.~A.,  {Johnston} K.~V.,  {Gandhi} S.~S.,
  {Laporte} C. F.~P.,   {B{\'e}dorf} J.,  2021, \mn@doi [\mnras]
  {10.1093/mnras/stab2580}, \href
  {https://ui.adsabs.harvard.edu/abs/2021MNRAS.508.1459H} {508, 1459}

\bibitem[\protect\citeauthoryear{{Jackson} et~al.,}{{Jackson}
  et~al.}{2021}]{Jackson2021}
{Jackson} R.~A.,  et~al., 2021, \mn@doi [\mnras] {10.1093/mnras/stab093}, \href
  {https://ui.adsabs.harvard.edu/abs/2021MNRAS.502.1785J} {502, 1785}

\bibitem[\protect\citeauthoryear{{Khoperskov}, {Di Matteo}, {Gerhard}, {Katz},
  {Haywood}, {Combes}, {Berczik}  \& {Gomez}}{{Khoperskov}
  et~al.}{2019}]{khoperskov2019buckling}
{Khoperskov} S.,  {Di Matteo} P.,  {Gerhard} O.,  {Katz} D.,  {Haywood} M.,
  {Combes} F.,  {Berczik} P.,   {Gomez} A.,  2019, \mn@doi [\aap]
  {10.1051/0004-6361/201834707}, \href
  {https://ui.adsabs.harvard.edu/abs/2019A&A...622L...6K} {622, L6}

\bibitem[\protect\citeauthoryear{{Kravtsov}, {Klypin}  \&
  {Khokhlov}}{{Kravtsov} et~al.}{1997}]{kr97ART}
{Kravtsov} A.~V.,  {Klypin} A.~A.,   {Khokhlov} A.~M.,  1997, \mn@doi [\apjs]
  {10.1086/313015}, \href
  {https://ui.adsabs.harvard.edu/abs/1997ApJS..111...73K} {111, 73}

\bibitem[\protect\citeauthoryear{{Lallement}, {Babusiaux}, {Vergely}, {Katz},
  {Arenou}, {Valette}, {Hottier}  \& {Capitanio}}{{Lallement}
  et~al.}{2019}]{Lallement2019}
{Lallement} R.,  {Babusiaux} C.,  {Vergely} J.~L.,  {Katz} D.,  {Arenou} F.,
  {Valette} B.,  {Hottier} C.,   {Capitanio} L.,  2019, \mn@doi [\aap]
  {10.1051/0004-6361/201834695}, \href
  {https://ui.adsabs.harvard.edu/abs/2019A&A...625A.135L} {625, A135}

\bibitem[\protect\citeauthoryear{{Laporte}, {Minchev}, {Johnston}  \&
  {G{\'o}mez}}{{Laporte} et~al.}{2019}]{laporte2019footprints}
{Laporte} C. F.~P.,  {Minchev} I.,  {Johnston} K.~V.,   {G{\'o}mez} F.~A.,
  2019, \mn@doi [\mnras] {10.1093/mnras/stz583}, \href
  {https://ui.adsabs.harvard.edu/abs/2019MNRAS.485.3134L} {485, 3134}

\bibitem[\protect\citeauthoryear{{Law} \& {Majewski}}{{Law} \&
  {Majewski}}{2010}]{law2010sagittarius}
{Law} D.~R.,  {Majewski} S.~R.,  2010, \mn@doi [\apj]
  {10.1088/0004-637X/714/1/229}, \href
  {https://ui.adsabs.harvard.edu/abs/2010ApJ...714..229L} {714, 229}

\bibitem[\protect\citeauthoryear{{Li} \& {Shen}}{{Li} \& {Shen}}{2020}]{Li2020}
{Li} Z.-Y.,  {Shen} J.,  2020, \mn@doi [\apj] {10.3847/1538-4357/ab6b21}, \href
  {https://ui.adsabs.harvard.edu/abs/2020ApJ...890...85L} {890, 85}

\bibitem[\protect\citeauthoryear{{Niederste-Ostholt}, {Belokurov}, {Evans}  \&
  {Pe{\~n}arrubia}}{{Niederste-Ostholt} et~al.}{2010}]{NiedersteOstholt2010}
{Niederste-Ostholt} M.,  {Belokurov} V.,  {Evans} N.~W.,   {Pe{\~n}arrubia} J.,
   2010, \mn@doi [\apj] {10.1088/0004-637X/712/1/516}, \href
  {https://ui.adsabs.harvard.edu/abs/2010ApJ...712..516N} {712, 516}

\bibitem[\protect\citeauthoryear{{Purcell}, {Bullock}, {Tollerud}, {Rocha}  \&
  {Chakrabarti}}{{Purcell} et~al.}{2011}]{purcell2011sagittarius}
{Purcell} C.~W.,  {Bullock} J.~S.,  {Tollerud} E.~J.,  {Rocha} M.,
  {Chakrabarti} S.,  2011, \mn@doi [\nat] {10.1038/nature10417}, \href
  {https://ui.adsabs.harvard.edu/abs/2011Natur.477..301P} {477, 301}

\bibitem[\protect\citeauthoryear{{Ramos}, {Antoja}  \& {Figueras}}{{Ramos}
  et~al.}{2018}]{ramos2018riding}
{Ramos} P.,  {Antoja} T.,   {Figueras} F.,  2018, \mn@doi [\aap]
  {10.1051/0004-6361/201833494}, \href
  {https://ui.adsabs.harvard.edu/abs/2018A&A...619A..72R} {619, A72}

\bibitem[\protect\citeauthoryear{{Roca-F{\`a}brega}, {Valenzuela},
  {Col{\'\i}n}, {Figueras}, {Krongold}, {Vel{\'a}zquez}, {Avila-Reese}  \&
  {Ibarra-Medel}}{{Roca-F{\`a}brega} et~al.}{2016}]{garrotxa}
{Roca-F{\`a}brega} S.,  {Valenzuela} O.,  {Col{\'\i}n} P.,  {Figueras} F.,
  {Krongold} Y.,  {Vel{\'a}zquez} H.,  {Avila-Reese} V.,   {Ibarra-Medel} H.,
  2016, \mn@doi [\apj] {10.3847/0004-637X/824/2/94}, \href
  {https://ui.adsabs.harvard.edu/abs/2016ApJ...824...94R} {824, 94}

\bibitem[\protect\citeauthoryear{{Roca-F{\`a}brega} et~al.,}{{Roca-F{\`a}brega}
  et~al.}{2021}]{RocaFabrega2021}
{Roca-F{\`a}brega} S.,  et~al., 2021, \mn@doi [\apj]
  {10.3847/1538-4357/ac088a}, \href
  {https://ui.adsabs.harvard.edu/abs/2021ApJ...917...64R} {917, 64}

\bibitem[\protect\citeauthoryear{{Rodr{\'\i}guez-Puebla}, {Primack},
  {Avila-Reese}  \& {Faber}}{{Rodr{\'\i}guez-Puebla}
  et~al.}{2017}]{rodriguez2017constraining}
{Rodr{\'\i}guez-Puebla} A.,  {Primack} J.~R.,  {Avila-Reese} V.,   {Faber}
  S.~M.,  2017, \mn@doi [\mnras] {10.1093/mnras/stx1172}, \href
  {https://ui.adsabs.harvard.edu/abs/2017MNRAS.470..651R} {470, 651}

\bibitem[\protect\citeauthoryear{{Ruiz-Lara}, {Gallart}, {Bernard}  \&
  {Cassisi}}{{Ruiz-Lara} et~al.}{2020}]{RuizLara2020}
{Ruiz-Lara} T.,  {Gallart} C.,  {Bernard} E.~J.,   {Cassisi} S.,  2020, \mn@doi
  [Nature Astronomy] {10.1038/s41550-020-1097-0}, \href
  {https://ui.adsabs.harvard.edu/abs/2020NatAs...4..965R} {4, 965}

\bibitem[\protect\citeauthoryear{{Sellwood}}{{Sellwood}}{2012}]{sellwood2012spiral}
{Sellwood} J.~A.,  2012, \mn@doi [\apj] {10.1088/0004-637X/751/1/44}, \href
  {https://ui.adsabs.harvard.edu/abs/2012ApJ...751...44S} {751, 44}

\bibitem[\protect\citeauthoryear{{Tian}, {Liu}, {Wu}, {Xiang}  \&
  {Zhang}}{{Tian} et~al.}{2018}]{Tian2018}
{Tian} H.-J.,  {Liu} C.,  {Wu} Y.,  {Xiang} M.-S.,   {Zhang} Y.,  2018, \mn@doi
  [\apjl] {10.3847/2041-8213/aae1f3}, \href
  {https://ui.adsabs.harvard.edu/abs/2018ApJ...865L..19T} {865, L19}

\bibitem[\protect\citeauthoryear{{Turk}, {Smith}, {Oishi}, {Skory}, {Skillman},
  {Abel}  \& {Norman}}{{Turk} et~al.}{2011}]{yt}
{Turk} M.~J.,  {Smith} B.~D.,  {Oishi} J.~S.,  {Skory} S.,  {Skillman} S.~W.,
  {Abel} T.,   {Norman} M.~L.,  2011, \mn@doi [The Astrophysical Journal
  Supplement Series] {10.1088/0067-0049/192/1/9}, \href
  {http://adsabs.harvard.edu/abs/2011ApJS..192....9T} {192, 9}

\bibitem[\protect\citeauthoryear{{Vasiliev} \& {Belokurov}}{{Vasiliev} \&
  {Belokurov}}{2020}]{Vasiliev2020}
{Vasiliev} E.,  {Belokurov} V.,  2020, \mn@doi [\mnras]
  {10.1093/mnras/staa2114}, \href
  {https://ui.adsabs.harvard.edu/abs/2020MNRAS.497.4162V} {497, 4162}

\bibitem[\protect\citeauthoryear{{Weinberg}}{{Weinberg}}{1989}]{Weinberg1989}
{Weinberg} M.~D.,  1989, \mn@doi [\mnras] {10.1093/mnras/239.2.549}, \href
  {https://ui.adsabs.harvard.edu/abs/1989MNRAS.239..549W} {239, 549}

\bibitem[\protect\citeauthoryear{{Weinberg}}{{Weinberg}}{1998}]{Weinberg1998}
{Weinberg} M.~D.,  1998, \mn@doi [\mnras] {10.1046/j.1365-8711.1998.01790.x},
  \href {https://ui.adsabs.harvard.edu/abs/1998MNRAS.299..499W} {299, 499}

\bibitem[\protect\citeauthoryear{{Widmark}, {Laporte}, {de Salas}  \&
  {Monari}}{{Widmark} et~al.}{2021}]{widmark2021weighing}
{Widmark} A.,  {Laporte} C.~F.~P.,  {de Salas} P.~F.,   {Monari} G.,  2021,
  \mn@doi [\aap] {10.1051/0004-6361/202141466}, \href
  {https://ui.adsabs.harvard.edu/abs/2021A&A...653A..86W} {653, A86}

\bibitem[\protect\citeauthoryear{{Widrow}, {Gardner}, {Yanny}, {Dodelson}  \&
  {Chen}}{{Widrow} et~al.}{2012}]{widrow2012galactoseismology}
{Widrow} L.~M.,  {Gardner} S.,  {Yanny} B.,  {Dodelson} S.,   {Chen} H.-Y.,
  2012, \mn@doi [\apjl] {10.1088/2041-8205/750/2/L41}, \href
  {https://ui.adsabs.harvard.edu/abs/2012ApJ...750L..41W} {750, L41}

\bibitem[\protect\citeauthoryear{{Williams} et~al.,}{{Williams}
  et~al.}{2013}]{williams2013wobbly}
{Williams} M.~E.~K.,  et~al., 2013, \mn@doi [\mnras] {10.1093/mnras/stt1522},
  \href {https://ui.adsabs.harvard.edu/abs/2013MNRAS.436..101W} {436, 101}

\makeatother
\end{thebibliography}







\bsp	
\label{lastpage}
\end{document}